\documentclass[11pt,a4paper]{article}

\usepackage[a4paper,margin=1in]{geometry}
\usepackage{natbib}
\usepackage[T1]{fontenc}
\usepackage[utf8]{inputenc}
\usepackage{microtype}

\usepackage{amsmath,amssymb,amsfonts,amsthm}
\usepackage{mathtools}
\usepackage{bm}

\usepackage{graphicx}
\usepackage{booktabs}
\usepackage{array}
\usepackage{multirow}
\usepackage{caption}
\usepackage{subcaption}
\usepackage[colorlinks=true,
            linkcolor=blue,
            citecolor=blue,
            urlcolor=blue]{hyperref}

\usepackage{enumitem}
\setlist{nosep}

\theoremstyle{plain}
\newtheorem{theorem}{Theorem}[section]
\newtheorem{lemma}[theorem]{Lemma}

\theoremstyle{definition}

\theoremstyle{remark}
\newtheorem{remark}[theorem]{Remark}

\newtheorem{assumption}[theorem]{Assumption}

\newcommand{\proper}{\mathsf}

\newcommand{\scal}[2]{\left\langle {#1},\,{#2} \right\rangle}

\newcommand{\E}{{\bf \proper{E}}}

\usepackage[normalem]{ulem}

\newcommand{\grad}{{\nabla}}
\newcommand{\calC}{{\cal C}}

\newcommand{\calH}{{\cal H}}

\newcommand{\calI}{{\cal I}}
\newcommand{\calJ}{{\cal J}}

\newcommand{\calN}{{\cal N}}

\let\oldhref\href
\renewcommand{\href}[2]{\oldhref{#1}{\bfseries#2}}

\newcommand{{\pCN}}{{$\text{pCN}$ }}
\newcommand{{\pCNL}}{{$\text{pCNL}$ }}
\newcommand{{\pCNAM}}{{$\text{pCN}_{\text{AM}}$ }}
\newcommand{{\pCNAP}}{{$\text{pCN}_{\text{AP}}$ }}
\newcommand{{\pCNAMo}}{{$\text{pCN}_{\text{AM}_0}$ }}
\newcommand{{\pCNLAM}}{{$\text{pCNL}_{\text{AM}}$ }}
\newcommand{{\pCNLHM}}{{$\text{pCNL}_{\text{HM}}$ }}
\newcommand{{\pCNLAP}}{{$\text{pCNL}_{\text{AP}}$ }}

\usepackage{comment}

\title{\bfseries Mesh Invariant Infinite Dimensional Adaptive MCMC for Latent Gaussian Processes}

\author{
Jonas Wallin\thanks{
  Department of statistics,
  Lund University,
  Sweden.
Email: jonas@stat.lu.se}
\and
Sreekar VAdlamani\thanks{TIFR--CAM, Bangalore, India. Email: sreekar@tifrbng.res.in}
}

\date{\today}

\begin{document}

\maketitle

\begin{abstract}
We introduce mesh-invariant adaptive Markov chain Monte Carlo methods for Gaussian-process posteriors arising in infinite-dimensional Bayesian inference. In function-space MCMC, posterior distributions are defined by a change of measure with respect to a Gaussian prior, making absolute continuity essential for valid proposal construction. Standard adaptive schemes that modify means or scales in the full discretized space can destroy this property, leading to proposal measures that become singular in the infinite-dimensional limit. To avoid this, we adapt only an active finite-dimensional subspace of the Gaussian-process representation while preserving the prior dynamics on inactive coordinates. This yields two adaptive proposals, pCNLV and pCNMV, which extend preconditioned Crank--Nicolson and Crank--Nicolson Langevin methods by learning posterior scale, and in pCNMV also posterior mean structure, on the data-informed subspace without introducing discretization-dependent Gaussian density ratios. The resulting samplers retain the mesh robustness of function-space methods while improving efficiency through local adaptation. Experiments on a Darcy-flow inverse problem and Bayesian logistic regression demonstrate consistent efficiency gains, including an approximately fourfold improvement in effective sampling efficiency for Darcy flow.
\end{abstract}


\section{Introduction}
\label{sec:intro}

Latent Gaussian process models are widely used in Bayesian inverse problems,
spatial statistics, data assimilation, and machine learning. When the unknown is
a function, field, or path, the posterior is naturally defined on an
infinite-dimensional Hilbert space. In this regime, standard finite-dimensional
MCMC methods typically deteriorate under mesh refinement, motivating algorithms
formulated directly on function space; see, for example,
\cite{stuart2004pathsampling,hairer2007spde-path-sampling,beskos2008diffbridges,
cotter2013mcmc,cui2016dili-mcmc,cui2024multilevel-dili,kazashi2026mgmc}.
Recent work on multilevel, parallel, neural-operator-accelerated, and
wide-neural-network samplers further shows the continuing importance of
mesh-stable MCMC methods
\cite{glattholtz2024parallel,cao2025dino,pezzetti2025functionspace}.

The function-space samplers of \cite{cotter2013mcmc} are built from
Crank--Nicolson discretizations of stochastic dynamics that preserve the
Gaussian reference measure. This construction gives the preconditioned
Crank--Nicolson sampler (pCN), the function-space analogue of random-walk
Metropolis, and the preconditioned Crank--Nicolson Langevin sampler (pCNL), the
corresponding Langevin analogue \cite{grenander1994representations}. These
methods are robust with respect to discretization dimension, but their efficiency
can be poor when the posterior mean, scale, or likelihood-informed directions
differ substantially from the prior.

A large body of work therefore incorporates posterior information into
function-space proposals. Examples include operator-weighted proposals
\cite{law2014speedupmcmc}, generalized pCN proposals \cite{rudolf2018gpcn},
adaptive pCN methods \cite{hu2017adaptivepcn}, adaptive independence samplers
\cite{feng2017adaptiveinfdim}, hybrid adaptive schemes
\cite{zhou2017hybridadaptive}, likelihood-informed subspace methods
\cite{cui2016dili-mcmc}, and geometric MCMC methods
\cite{beskos2017geometric,cao2025dino}. These approaches can capture richer
posterior geometry, including correlations between prior modes, but typically
require gradient, Hessian, subspace, or other problem-specific information.

We propose a simpler adaptive alternative based on changing the Gaussian
reference measure while preserving equivalence to the prior. Let
\(\mu_0=\calN(0,\calC)\) be the prior on \(\calH\), and suppose that
\begin{align}
\label{eqn:prior-posterior}
\frac{d\mu_1}{d\mu_0}(u)=Z^{-1}\exp\bigl(-\Phi(u;Y)\bigr),
\end{align}
where $Z>0$ is the normalizing constant
For any Gaussian measure \(\Psi\) equivalent to \(\mu_0\), the same posterior
can be written as
\[
\frac{d\mu_1}{d\Psi}(u)\propto
\exp\bigl(-\widetilde{\Phi}(u;Y)\bigr),
\qquad
\widetilde{\Phi}(u;Y)
=
\Phi(u;Y)+\log\!\left(\frac{d\Psi}{d\mu_0}(u)\right).
\]
This change-of-reference viewpoint lets us adapt the proposal to the posterior
while avoiding discretization-dependent Gaussian density ratios.

Our contributions are as follows. First, we derive two mesh-stable adaptive
samplers from equivalent Gaussian changes of reference: a variance-adapted pCNL
proposal, \(\text{pCNL}_{\text{V}}\), and a mean--variance-adapted pCN proposal,
\(\text{pCN}_{\text{MV}}\). Second, we give practical rules for learning the adapted
mean and scale parameters, based on Kullback--Leibler  matching and a diagonal inverse-Fisher
scaling motivated by \citet{titsias2023fisher}. Third, we introduce an online
truncated adaptation scheme that updates only an active finite-dimensional set
of prior modes and leaves the remaining modes unchanged. This truncation is what
preserves mesh invariance: the adapted Gaussian reference remains equivalent to
the prior at every iteration, so the sampler stays well defined in the
infinite-dimensional limit rather than becoming a discretization-dependent
adaptive method.

Empirically, on the nonlinear Darcy flow inverse problem of
\citet{wang2026global}, the gradient-free \(\text{pCN}_{\text{MV}}\)
sampler gives more than a four-fold improvement in effective sample size over
adaptive pCN while recovering the dominant posterior conductivity structure.
On Bayesian logistic-regression benchmarks following
\citet{titsias2016auxiliary,titsias2023fisher}, the adapted pCNL and
\(\text{pCN}_{\text{MV}}\) variants consistently improve  Effective sample size per second
over the corresponding pCN and pCNL baselines, with the best method depending on
whether gradient information is computationally worthwhile.

The price of this simplicity is that the adaptation is diagonal in the prior basis. Thus the method requires knowledge of the prior basis and a useful ordering of its modes, and it does not learn correlations between different prior directions. When such correlations dominate, likelihood-informed, hybrid adaptive, low-rank, or geometric methods may be more effective \cite{spantini2015optimal,cui2016dili-mcmc,cao2025dino}. Our methods are therefore complementary: they provide a lightweight gradient-free or low-gradient alternative when full posterior-geometry adaptation is too expensive.

{\it Organization of the paper.} Section~\ref{sec:background} reviews Gaussian
measures on Hilbert spaces and the notation used throughout.
Section~\ref{sec:proposals-results} introduces the proposed samplers.
Section~\ref{sec:estimation} discusses online estimation of the adaptive
quantities, and Section~\ref{sec:experiments} reports numerical results.


\section{Background}
\label{sec:background}

\subsection{Gaussian measures in infinite dimensions}
\label{subsec:gauss-meas}

Throughout the paper, \(\calH\) denotes a separable Hilbert space with inner product  \(\scal{\cdot}{\cdot}\) and norm \(\|\cdot\|\). We consider a centered Gaussian prior measure \(\mu_0=\calN(0,\calC)\) on \(\calH\), where the covariance operator \(\calC:\calH\to\calH\) is self-adjoint, positive, injective, and trace class. By the spectral theorem for compact self-adjoint operators, \(\calC\) admits an orthonormal eigenbasis; see, for example, Section 2 of \cite{stuart2010inverse}.
\vspace{-0.1cm}
\begin{assumption}[Spectral structure of the prior covariance]
\label{ass:spectral-structure}
There exists an orthonormal basis \(\{e_k\}_{k\ge 1}\) of \(\calH\) and a non-increasing sequence of positive eigenvalues \(\{\sigma_k\}_{k\ge 1}\) with \(\sigma_k\downarrow 0\) such that \(\calC e_k=\sigma_k e_k\) for every \(k\). Equivalently,
$
\calC=\sum_{k\ge 1}\sigma_k \scal{\cdot}{e_k}e_k.
$
In what follows, \(\{(\sigma_k,e_k)\}_{k\ge 1}\) denotes the ordered eigensystem of \(\calC\).
\end{assumption}
\vspace{-0.1cm}
It is convenient to represent \(\calC\) through its diagonal form in the eigenbasis. Let \(S:\calH\to \ell^2\) be the unitary coordinate map defined by \(Su=(\scal{u}{e_k})_{k\ge 1}\), and let \(\Lambda:\ell^2\to\ell^2\) be the diagonal operator given by \(\Lambda(x_1,x_2,\dots)=(\sigma_1x_1,\sigma_2x_2,\dots)\). Then
\begin{equation}
\label{eq:cov-diagonal}
\calC=S^*\Lambda S.
\end{equation}
Thus, if \(u=\sum_{k\ge 1}u_k e_k\) with \(u_k=\scal{u}{e_k}\), then \(\calC\) acts diagonally as \(\calC u=\sum_{k\ge 1}\sigma_k u_k e_k\). This representation will be used throughout the paper to define adapted proposals and their finite-dimensional approximations.

Under Assumption~\ref{ass:spectral-structure}, a draw \(u\sim\mu_0\) admits the Karhunen--Lo\`eve expansion
$
u=\sum_{k\ge 1}\sigma_k^{1/2}\xi_k e_k,
$
where \((\xi_k)_{k\ge 1}\) are i.i.d.\ standard Gaussian random variables, with convergence in \(\calH\) almost surely and in \(L^2\).

The covariance operator also induces the Cameron--Martin geometry. We define the Cameron--Martin space by \(\calH_{\calC}=\operatorname{Im}(\calC^{1/2})\), endowed with inner product \(\scal{x}{y}_{\calC}=\scal{\calC^{-1/2}x}{\calC^{-1/2}y}\) and norm \(\|x\|_{\calC}=\|\calC^{-1/2}x\|\). Equivalently, if \(x=\sum_{k\ge 1}x_k e_k\), then \(x\in\calH_{\calC}\) if and only if \(\sum_{k\ge 1} x_k^2/\sigma_k<\infty\), in which case \(\|x\|_{\calC}^2=\sum_{k\ge 1}x_k^2/\sigma_k\).

The following classical result characterizes equivalence of Gaussian measures on Hilbert spaces; see, for example, Theorem 3.4 in \cite{kuo1970gaussian}.

\begin{theorem}[Feldman--Hajek]
\label{thm:feldman-hajek}
Let \(\calN(m_1,\calC_1)\) and \(\calN(m_2,\calC_2)\) be Gaussian measures on \(\calH\). Then they are either equivalent or mutually singular. They are equivalent if and only if \(m_1-m_2\in\calH_{\calC_1}\) and the operator \(\left(\calC_1^{-1/2}\calC_2\calC_1^{-1/2}-\calI\right)\) is Hilbert--Schmidt.
\end{theorem}


\subsection{Overview of MCMC methods in infinite dimensions}
\label{subsec:overview-mcmc}

Our interest lies in MCMC methods for sampling from the posterior measure \(\mu_1\) on \(\calH\). Throughout this section, \(\mu_1\) is assumed to satisfy the change-of-measure relation \eqref{eqn:prior-posterior} with respect to the centered Gaussian priorprior measure \(\mu_0=\calN(0,\calC)\).

A convenient starting point for the preconditioned proposals considered here is the SPDE
\begin{equation}
\label{eqn:spde}
\frac{\partial u}{\partial s}
=
-u-\gamma \calC\grad\Phi(u)
+\sqrt{2\calC}\,\frac{dB}{ds},
\end{equation}
where \(B\) is a cylindrical Brownian motion on \(\calH\), and \(\gamma\in\{0,1\}\). When \(\gamma=0\), \eqref{eqn:spde} reduces to an Ornstein--Uhlenbeck dynamics with invariant measure \(\mu_0\). When \(\gamma=1\), and under standard regularity assumptions on \(\Phi\), the invariant measure is \(\mu_1\). We note that, throughout the subsequent analysis, we assume that $\Phi$ satisfies Assumptions 6.1 in \cite{cotter2013mcmc}. We shall invoke additional assumptions later, as and when needed. The pCN and pCNL proposals are obtained by Crank--Nicolson discretizations of these two cases.

Once a proposal kernel \(q(u,dv)\) is specified, we define the joint proposal-target measure
$
\nu(du,dv)=q(u,dv)\,\mu_1(du),
$
and its reversal
$
\nu^\perp(du,dv)=q(v,du)\,\mu_1(dv).
$
Whenever \(\nu^\perp\) is absolutely continuous with respect to \(\nu\), we write
$
\frac{d\nu^\perp}{d\nu}(u,v)=\exp\bigl(\calJ(u,v)\bigr).
$
The corresponding Metropolis--Hastings acceptance probability is then
\begin{equation}
\label{eqn:accept-reject}
\alpha(u,v)=\min\Bigl(1,\exp\bigl(\calJ(u,v)\bigr)\Bigr).
\end{equation}
Thus, for each proposal it suffices to identify the corresponding acceptance correction \(\calJ(u,v)\).

\subsubsection{Preconditioned Crank--Nicolson proposal (pCN)}
\label{subsubsec:pCN}

The preconditioned Crank--Nicolson proposal is obtained by applying a Crank--Nicolson discretization to \eqref{eqn:spde} with \(\gamma=0\). Since the underlying dynamics preserve the Gaussian base measure \(\mu_0\), the resulting proposal uses only the prior geometry encoded by \(\calC\). This exact preservation of the reference Gaussian structure is what makes pCN well behaved in infinite dimensions and robust under mesh refinement.

The proposal takes the autoregressive form
\begin{equation}
\label{eqn:pCN-proposal}
v=(1-\beta^2)^{1/2}u+\beta \calC^{1/2}\xi,
\end{equation}
where \(\xi\sim\calN(0,\calI)\) and \(\beta\in(0,1]\). Because this proposal is reversible with respect to \(\mu_0\), the Metropolis--Hastings correction depends only on the change in the potential 
$
\calJ_{\text{pCN}}(u,v)=\Phi(u)-\Phi(v).
$

\subsubsection{Preconditioned Crank--Nicolson Langevin proposal (pCNL)}
\label{subsubsec:pCNL}

The preconditioned Crank--Nicolson Langevin proposal is obtained from the same discretization of \eqref{eqn:spde}, now with \(\gamma=1\). Unlike pCN, it incorporates local first-order information from the target through \(\grad\Phi(u)\). The proposal still respects the covariance structure induced by \(\calC\), while the additional drift term biases the proposals to move toward regions of higher posterior probability. Let
$
c_\beta = 1-\sqrt{(1-\beta^2)},
$
then the proposal is
\begin{equation}
\label{eqn:pCNL-proposal}
v=(1-c_\beta) u-c_\beta \calC\grad\Phi(u)+\beta \calC^{1/2}\xi,
\end{equation}
where \(\xi\sim\calN(0,\calI)\) and \(\beta\in(0,1]\). 
The corresponding Metropolis--Hastings correction is more involved than for pCN, since the proposal is no longer reversible with respect to \(\mu_0\). In terms of \(\beta\), it is given by
{\small
\begin{align*}
\calJ_{\text{pCNL}}(u,v)
&= \Phi(u)-\Phi(v)
 + \frac12\scal{v-u}{\grad\Phi(u)+\grad\Phi(v)} + \frac{c_\beta^2}{2\beta^2}
\left[
\begin{aligned}
&\|\calC^{1/2}\grad\Phi(u)\|^2
 - \|\calC^{1/2}\grad\Phi(v)\|^2 \\
&\quad + \scal{u+v}{\grad\Phi(u)-\grad\Phi(v)}
\end{aligned}
\right].
\end{align*}}


\section{Variance- and mean-adapted proposals}
\label{sec:proposals-results}

In this section we introduce two adaptive proposals obtained by changing the Gaussian reference measure while keeping the target measure fixed. 
The first is a \emph{variance-adapted} proposal, which modifies the covariance structure of the reference measure so that proposals better reflect the local scale of the posterior, resulting in an adapted pCNL.
 The second is a \emph{mean-variance-adapted} proposal, which in addition shifts the proposal toward the posterior mean and provides a gradient-free alternative when derivatives of \(\Phi\) are unavailable,
 resulting in an adapted pCN. In both cases, the objective is to incorporate posterior information while preserving the structure needed for the resulting methods to remain well defined in infinite dimensions and computationally tractable. We first describe these proposals in the idealized setting where the relevant adaptation parameters are known, and later discuss how they can be estimated in practice.

Recall from \eqref{eq:cov-diagonal} that the prior covariance admits the representation
$
\calC = S^T \Lambda S,
$
where \(S\) is the orthonormal basis map associated with the eigenbasis of \(\calC\), and \(\Lambda=\operatorname{diag}(\sigma_1,\sigma_2,\ldots)\) contains the eigenvalues of \(\calC\). We consider adaptive modifications of the prior covariance of the form
\[
\calC_d = S^T \Lambda D S,
\qquad D=\operatorname{diag}(d_1,d_2,\ldots),
\]
with \(d_k>0\) for all \(k\). Thus the adapted covariance preserves the eigenfunctions of \(\calC\) and modifies only the scale of each mode.

\subsection{Changing the base measure}

The constructions in Section~\ref{subsec:overview-mcmc} depend on the choice of Gaussian base measure. We therefore replace the original prior \(\mu_0=\calN(0,\calC)\) by the adapted Gaussian reference
\[
\mu_{d,m}=\calN(m,\calC_d),
\qquad \calC_d = S^T \Lambda D S.
\]
We assume that \(\mu_{d,m}\) is equivalent to \(\mu_0\).\footnote{Since \(\calC\) and \(\calC_d\) share the same eigenvectors, equivalence requires the diagonal Feldman--Hajek condition \(\sum_{k\ge1}(d_k-1)^2<\infty\), together with \(m\) belonging to the common Cameron--Martin space.} The target measure \(\mu_1\) can then be re-expressed relative to \(\mu_{d,m}\) as
\[
\frac{d\mu_1}{d\mu_{d,m}}(u)
=
\frac{d\mu_1}{d\mu_0}(u)\,
\frac{d\mu_0}{d\mu_{d,m}}(u)
\propto
\exp\bigl(-\widetilde\Phi_{d,m}(u)\bigr),
\]
where, up to an additive constant, and using the diagonal representation
{\small
\begin{equation}
\label{eqn:adapt-potential-expanded}
\begin{aligned}
\widetilde\Phi_{d,m}(u)
={}&
\Phi(u)
+
\frac12\scal{u}{S^T\Lambda^{-1}(I-D^{-1})Su}
+
\scal{u}{S^T D^{-1}\Lambda^{-1}Sm}.
\end{aligned}
\end{equation}}

This change of reference measure leads naturally to a mean- and variance-adapted version of pCN and pCNL, obtained by replacing \(\calC\) by \(\calC_d\), \(\Phi\) by \(\widetilde\Phi_{d,m}\), and centering the proposal at \(m\).


\subsection{Mean-variance-adjusted pCN}
\label{subsec:pcn-mv}

Given the adjusted Gaussian reference \(\mu_{d,m}=\calN(m,\calC_d)\) and modified potential \(\widetilde\Phi_{d,m}\) from \eqref{eqn:adapt-potential-expanded}, we define the mean-variance-adjusted proposals by applying pCN and pCNL relative to \(\mu_{d,m}\).

For the mean-variance-adjusted pCN proposal, \(\text{pCN}_{\text{MV}}\) , given the current state \(u\), the proposal is
\begin{equation}
\label{eqn:pCN-MV-proposal}
v
=
(1-c_\beta) u + c_\beta m +\beta \calC_d^{1/2}\xi,
\end{equation}
where $\xi\sim\calN(0,\calI)$.
Since this proposal is reversible with respect to \(\calN(m,\calC_d)\), the Metropolis--Hastings correction is
\begin{equation}
\label{eqn:pCN-MV-J}
\calJ_{\text{pCN}_{\text{MV}}}(u,v)
=
\widetilde\Phi_{d,m}(u)-\widetilde\Phi_{d,m}(v).
\end{equation}

For the mean-variance-adjusted pCNL proposal, we use the gradient of the modified potential,
\[
\grad\widetilde\Phi_{d,m}(u)
=
\grad\Phi(u)
+
(\calC^{-1}-\calC_d^{-1})u
+
\calC_d^{-1}m,
\]
and proposals generated as
$
v
=(1-c_\beta) u + c_\beta \left(m- \calC_d \grad\widetilde\Phi_{d,m}(u) \right)
+\beta \calC_d^{1/2}\xi,
$
where $\xi\sim\calN(0,\calI)$.
We note that
$
m-\calC_d\grad\widetilde\Phi_{d,m}(u)
=
u-\calC_d\bigl(\calC^{-1}u+\grad\Phi(u)\bigr).
$
Hence the \(m\)-terms cancel in the proposal, and the mean of the reference measure has no effect on the resulting pCNL dynamics. For this reason, in the Langevin case it is natural to work directly with the centered adapted reference \(\mu_{0,d}=\calN(0,\calC_d)\) and treat only the covariance multipliers \(D\) as adaptive parameters.

The mean parameter \(m\) cancels from the Langevin proposal once the modified potential is substituted, so the pCNL adaptation depends only on the covariance multipliers \(D\). In contrast, \(m\) remains relevant for \(\text{pCN}_{\text{MV}}\), where it centers the autoregressive proposal. Moreover, since
$
\E_{\mu_1}\!\left[\grad\widetilde\Phi_{d,m}(u)\right]=0,
$
\(\text{pCN}_{\text{MV}}\) may be interpreted as a gradient-free version of mean variance adjusted  pCNL, obtained by replacing the local drift \(\grad\widetilde\Phi_{d,m}(u)\) by its posterior expectation.


\subsection{Variance-adjusted pCNL}
\label{subsec:pcnl-v}

For pCNL, if we study the calculations the $m$ terms cancels. This motivates a simpler variant in which only the covariance is adjusted, while the reference measure remains centered. Let \(\mu_{0,d}=\calN(0,\calC_d)\) with \(\calC_d=S^T\Lambda D S\).
Given the current state \(u\), we propose
\begin{equation}
\label{eqn:pCNL-V-proposal}
v
=
(1-c_\beta) u
- c_\beta \calC_d \grad \widetilde\Phi_d(u)
+
\beta \calC_d^{1/2}\xi,
\end{equation}
with $\xi\sim\calN(0,\calI).$

The corresponding Metropolis--Hastings correction is obtained from the pCNL formula with \(\Phi\) replaced by \(\widetilde\Phi_d\) and \(\calC\) replaced by \(\calC_d\):
{\footnotesize
\begin{align*}
\calJ_{\text{pCNL}_{\text{V}}}(u,v)
= & \widetilde\Phi_d(u)-\widetilde\Phi_d(v)+  \frac12\scal{v-u}{\grad\widetilde\Phi_d(u)+\grad\widetilde\Phi_d(v)}
\\
&
+\frac{c_\beta^2}{2\beta^2}
\Bigl(
\|\calC_d^{1/2}\grad\widetilde\Phi_d(u)\|^2
-\|\calC_d^{1/2}\grad\widetilde\Phi_d(v)\|^2
+\scal{u+v}{\grad\widetilde\Phi_d(u)-\grad\widetilde\Phi_d(v)}
\Bigr).
\end{align*}
}
A useful connection to the marginal gradient sampler of \citet[Eq.~(8)]{titsias2016auxiliary} is that, with step-size parameter $\delta$ as defined there, its proposal induces a mode-wise rescaling of the covariance eigenvalues; see also the spectral discussion in Section~3.4 of \citet{titsias2016auxiliary}. In our notation, the same covariance scaling is recovered by the parametric choice
$
d_k=\frac{\delta(\delta+4\sigma_k)}{(\delta+2\sigma_k)^2},
$
where $\sigma_k$ denotes the $k$th eigenvalue of $\calC$. The two samplers are nevertheless not identical, since their proposal means are different.


\subsection{Kullback--Leibler characterization of the adaptation}
\label{subsec:KL}

The constructions in Section~\ref{sec:proposals-results} leaves open how the adaptive quantities \(D\) and \(m\) should be chosen. A natural and principled answer is obtained by approximating the target measure \(\mu_1\) with a Gaussian measure that is optimal in Kullback--Leibler divergence. This provides a direct interpretation of the proposals introduced above: the variance-adapted pCNL proposal uses the covariance of the KL-optimal Gaussian approximation, while the mean-variance-adapted pCN proposal additionally uses the posterior mean.

Following \cite{stuart2015approxKLalgo,stuart2015approxKL,feng2017adaptiveinfdim}, we consider Gaussian measures of the form
$
\mu_{d,m}=\calN(m,\calC_d)$ , where
$\calC_d=S^T\Lambda D S,
$
 \(m\in\calH_{\calC}\) and \(D=\operatorname{diag}(d_1,d_2,\ldots)\) with \(d_k>0\). We restrict attention to Gaussian measures \(\Psi\) that are equivalent to \(\mu_0\), which in the present diagonal setting is ensured by
$
\sum_{k\ge1}(d_k-1)^2<\infty.
$
The reverse Kullback--Leibler divergence is
{\small
\[
\delta(\mu_1\|\mu_{d,m})
=
\int \log\Bigl(\frac{d\mu_1}{d\mu_{d,m}}(u)\Bigr)\,d\mu_1(u).
\]}
Since \(d\mu_1/d\mu_0 \propto \exp(-\Phi)\), minimizing \(\delta(\mu_1\|\mu_{d,m})\) over \(m\) and \(D\) is equivalent to maximizing \(\E_{\mu_1}[\log(d\mu_{d,m}/d\mu_0)(u)]\), or, equivalently, minimizing the Gaussian approximation term
$
-\E_{\mu_1}\Bigl[\log\frac{d\mu_{d,m}}{d\mu_0}(u)\Bigr].
$
Writing \((Su)_k\) and \((Sm)_k\) for the coordinates of \(u\) and \(m\) in the eigenbasis of \(\calC\), and using that \(\calC_d\) and \(\calC\) share the same eigenvectors, a straightforward calculation yields
{\small
\begin{equation}
\label{eqn:KL-objective}
\delta(\mu_1\|\mu_{d,m})
=
\text{const}
+
\frac{1}{2}
\sum_{k\ge1}
\log d_k
+
\frac{\E_{\mu_1}\bigl[((Su)_k-(Sm)_k)^2\bigr]}{\sigma_k d_k}
,
\end{equation}}
where the constant does not depend on \(m\) or \(D\). Hence the optimization decouples coordinatewise. Minimizing \eqref{eqn:KL-objective} over \(m\) and \(D\) gives
\begin{equation}
\label{eqn:opt-mean-variance}
m_\star=\E_{\mu_1}[u],
\qquad
d_{\star,k}
=
\frac{1}{\sigma_k}\,
\E_{\mu_1}\bigl[((Su)_k-(Sm_\star)_k)^2\bigr].
\end{equation}
Thus, the KL-optimal Gaussian approximation is obtained by matching the posterior mean and the posterior variances in the eigendirections of the prior covariance. In this sense, the adaptive quantities introduced in Section~\ref{sec:proposals-results} are naturally interpreted as KL-adapted parameters.

This characterization also clarifies the role of the two proposals introduced earlier. The variance-adapted pCNL proposal uses the covariance \(\calC_{d_\star}\) determined by \eqref{eqn:opt-mean-variance}, while retaining the pCNL structure based on the modified potential \(\widetilde\Phi_{d_\star}\). The mean-variance-adapted pCN proposal uses both \(m_\star\) and \(\calC_{d_\star}\), and may therefore be viewed as the proposal associated with the full KL-optimal Gaussian approximation \(\calN(m_\star,\calC_{d_\star})\).

To use \eqref{eqn:opt-mean-variance} in the infinite-dimensional setting, the minimizers must define a Gaussian measure equivalent to \(\mu_0\). This requires \(m_\star\in\calH_{\calC}\) and \(\sum_{k\ge1}(d_{\star,k}-1)^2<\infty\). The first condition follows from the Fisher identity \(\E_{\mu_1}[\grad\Phi(u)]=-\calC^{-1}m_\star\), together with the regularity assumptions on \(\grad\Phi\). The second is precisely the variance regularity condition imposed in Assumption (A3).

\begin{remark}
If the target measure itself is Gaussian and equivalent to \(\mu_0\), then the condition \(\sum_{k\ge1}(d_{\star,k}-1)^2<\infty\) is satisfied automatically. In case of non--Gaussian target measure, we assume that
\begin{equation}\label{eqn:A3}
\sum_{k\ge 1} \left( d_{\star,k}-1\right)^2 <\infty
\end{equation}
As can be seen from Theorem~\ref{thm:feldman-hajek}, the above assumption is a natural extension of the Gaussian equivalence condition to the non-Gaussian setting considered here.
\end{remark}



\subsection{A diagonal Fisher approximation for the covariance multipliers}
\label{subsec:fisher-d}

 In finite dimensions,
\citet{titsias2023fisher} show that, under a global scale constraint, the expected-squared-jump-distance
optimal preconditioner for preconditioned MALA is proportional to the inverse Fisher matrix.
Because in our setting the adapted covariance is restricted to the diagonal family
\(\calC_d=S^T\Lambda D S\), the corresponding function-space analogue is to retain only the
diagonal of the Fisher matrix in the prior Karhunen--Lo\`eve basis.

To make this precise, consider the finite dimensional case with mode whitened coordinates
$
\xi_k := (\Lambda^{-1/2}Su)_k
$
In these coordinates, the target density
satisfies
$
\pi_N(\xi)\propto \exp\Bigl(-\Phi(u)-\frac12\sum_{k=1}^N \xi_k^2\Bigr),
$
and therefore its score is given componentwise by
$
\partial_{\xi_k}\log \pi_N(\xi)
=
-\xi_k-\sigma_k^{1/2}(S\grad\Phi(u))_k.
$
, the proposal \eqref{eqn:pCNL-V-proposal} can be rewritten in these
coordinates as
$
\xi'
=
\xi+ (1-a_\beta) D_N \nabla_{\xi}\log \pi_N(\xi)+\beta D_N^{1/2}\zeta,
$
where \(D_N=\operatorname{diag}(d_1,\dots,d_N)\). Thus \(D_N\) plays exactly the role of a
diagonal preconditioner for a finite-dimensional Langevin proposal.

This suggests defining the diagonal Fisher quantities, and the corresponding scaling,
{\small
\begin{equation}
\label{eqn:fisher-dk}
\iota_k
:=
\E_{\mu_1}\!\left[
\left(
\sigma_k^{-1/2}(Su)_k+\sigma_k^{1/2}(S\grad\Phi(u))_k
\right)^2
\right],
\qquad
d_k^{\mathrm F}=\iota_k^{-1}.
\end{equation}}
Since the term inside the expectation is minus the \(k\)-th component of the whitened score,
\eqref{eqn:fisher-dk} is precisely the diagonal analogue of the inverse-Fisher preconditioner
advocated by \citet{titsias2023fisher}.


\section{Online estimation of the KL-adapted parameters}
\label{sec:estimation}

The proposals in Section~\ref{sec:proposals-results} were described in the idealized setting where the KL-adapted quantities are known. In practice, however, the optimal mean \(m_\star\) and variance multipliers \(d_{\star,k}\) from Section~\ref{subsec:KL} must be estimated from the MCMC output. These quantities play different roles in the two adaptive proposals: the variance-adapted pCNL proposal uses the covariance correction through \((d_{\star,k})\), while the mean-variance-adapted pCN proposal uses both the covariance correction and the posterior mean \(m_\star\). In this section we describe an online estimation scheme for these quantities that preserves the infinite-dimensional structure of the algorithm.
Let \(u^{(j)}\) denote the \(j\)-th state of the Markov chain, and define its
\(k\)-th whitened Karhunen--Loève coefficient by
$
\xi_k^{(j)} := (\Lambda^{-1/2}Su^{(j)})_k .
$
We estimate the posterior mean and second moment of these coefficients
recursively. For each \(k\), let
\[
\hat m_k^{(j)}=w_j\xi_k^{(j)}+(1-w_j)\hat m_k^{(j-1)}, 
\qquad
\hat q_k^{(j)}=w_j(\xi_k^{(j)})^2+(1-w_j)\hat q_k^{(j-1)} .
\]
with \(w_j = j^{-1}\). We then define the variance estimator by
$
\hat d_k^{(j)} := \hat q_k^{(j)} - \bigl(\hat m_k^{(j)}\bigr)^2.
$
These are the natural online estimators associated with the KL-optimal
quantities in \eqref{eqn:opt-mean-variance}.

A direct use of \(\hat m^{(j)}\) and \(\hat d^{(j)}\) does not, in general,
guarantee that the resulting Gaussian reference measure remains equivalent to
the prior. In particular, one would need the reconstructed mean
$
\hat m_{\mathrm{orig}}^{(j)} := S^\top \Lambda \hat m^{(j)}
$
to belong to \(\calH_{\calC}\), together with
\(\sum_{k\ge1}(\hat d_k^{(j)}-1)^2<\infty\), and these conditions need not hold for the raw online estimates. 

We, therefore, use a truncated adaptation
strategy described as follows: let \((N_j)_{j\ge1}\) be a slowly increasing sequence of positive integers. We define
\[
\tilde m_k^{(j)} =
\begin{cases}
\hat m_k^{(j)}, & k < N_j,\\
0, & k \ge N_j,
\end{cases}
\qquad
\tilde q_k^{(j)} =
\begin{cases}
\hat q_k^{(j)} + \epsilon, & k < N_j,\\
1, & k \ge N_j,
\end{cases}
\]
where $\epsilon>0$
and then set
$
\tilde d_k^{(j)} := \tilde q_k^{(j)} - \bigl(\tilde m_k^{(j)}\bigr)^2.
$
By construction,
$
\tilde m_{\mathrm{orig}}^{(j)} := S^\top \Lambda \tilde m^{(j)}
$
has only finitely many nonzero coefficients in the \(\calC\)-eigenbasis, and
\(\tilde d^{(j)}-1\) has finite support. 

\begin{lemma}
\label{lem:1}
Let $u^{(0)}\sim\mu_1$. Then, for any $j\ge 1$, we have that almost surely, \(\hat{m}_{\textrm{orig}}^{(j)} \notin \calH_{\calC}\), whereas 
\(\tilde m_{\mathrm{orig}}^{(j)} \in \calH_{\calC}\) almost surely. Also,
\(\sum_{k\ge1}(\tilde d_k^{(j)}-1)^2<\infty\) hold automatically at every iteration, so the adapted Gaussian reference remains equivalent to the prior.
\end{lemma}

The proof is given in Appendix \ref{sec:proof}.
In the numerical experiments below, we use schedules of the same form but tune the constants to the scale of each experiment. Specifically, we take
\[
N_j
=
N_{j-1}
+
K\,\mathbb{I}\bigl(j \;(\mathrm{mod}\; M)=0\bigr),
\]
where \(K\) is the number of additional coordinates introduced at each update and \(M\) is the update period. Thus, the adapted dimension grows gradually with the iteration count. In all experiments, \(K\) and \(M\) are chosen so that \(N_j\) increases slowly relative to the total number of MCMC iterations. More generally, \(N_j\) can be any increasing sequence of positive integers with sublinear growth.

The same truncated online-estimation strategy used for the KL-adapted parameters can also be applied to the diagonal Fisher approximation. We describe this construction in Appendix~\ref{sec:ONline}.


\section{Numerical experiments}
\label{sec:experiments}

We evaluate the proposed adaptation on three examples. The Gaussian
measurement-error benchmark provides an oracle setting in which \(m_\star\) and
\(d_\star\) are known exactly; the Darcy flow inverse problem assesses the
gradient-free method in a nonlinear latent-field model with an expensive forward
solve; and the Bayesian logistic-regression benchmarks test the adaptive
schemes on finite but moderately high-dimensional non-Gaussian posteriors.


\subsection{Gaussian benchmark and truncated adaptation}
\label{subsec:gaussian-benchmark}

\begin{figure}[t]
    \centering
    \includegraphics[width=0.3\linewidth,height=0.2\linewidth]{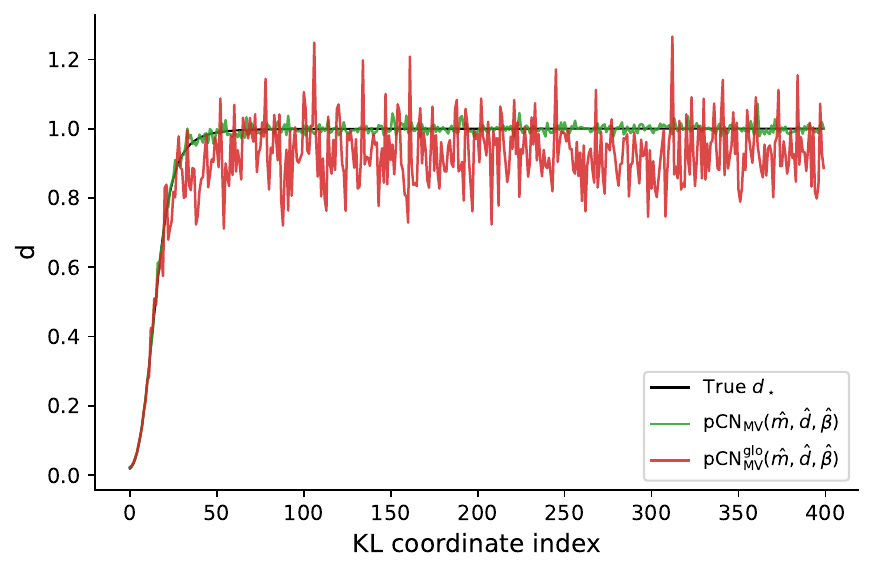}
    \includegraphics[width=0.3\linewidth,height=0.2\linewidth]{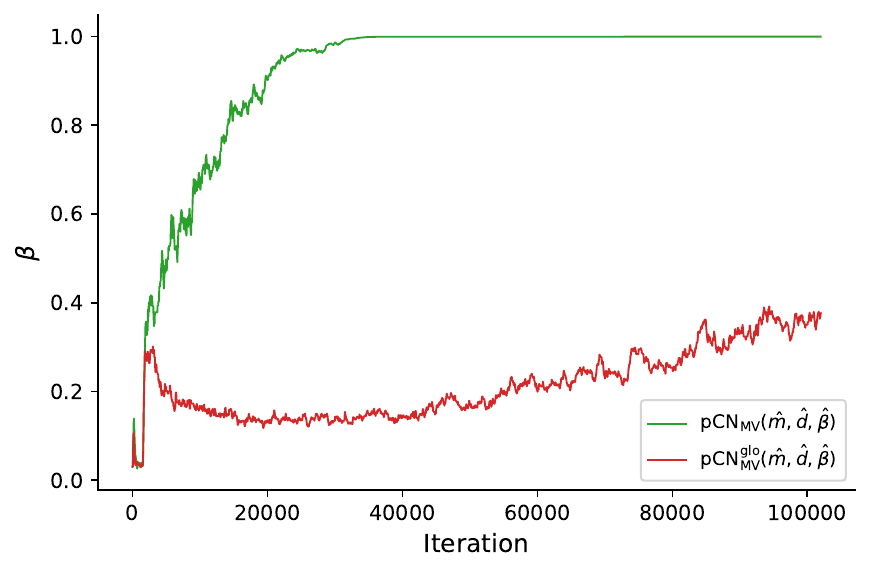}
        \includegraphics[width=0.3\linewidth,height=0.2\linewidth]{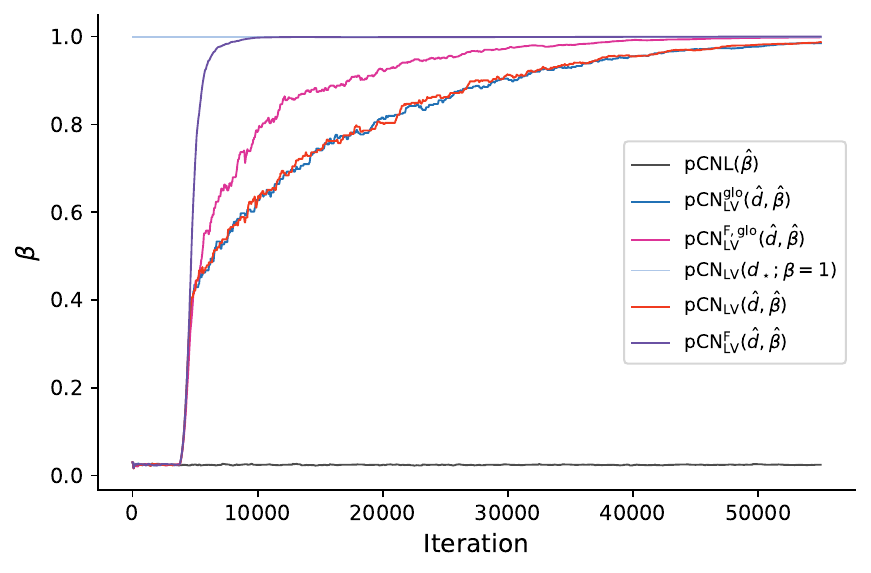}
\caption{
    Effect of truncating the online adaptation. Left: estimates of the variance
    multipliers \(d_k\) in KL coordinates for the truncated
    \(\mathrm{pCN}_{\mathrm{MV}}\) updates, and the oracle
    values \(d_{\star,k}\). Middle: adaptation of \(\beta\) for
    \(\mathrm{pCN}_{\mathrm{MV}}\) variants under target acceptance \(0.234\). Right:
    adaptation of \(\beta\) for \(\mathrm{pCNL}_{\mathrm V}\) variants under target
    acceptance \(0.53\).
    }
    \label{fig:pcnmv-adapt}
\end{figure}
We next use a conjugate Gaussian benchmark to isolate the effect of adaptation from approximation error in the adapted parameters. Let \(u\sim\calN(0,\calC)\) and \(y=u+\epsilon\), with \(\epsilon\sim\calN(0,\sigma_\epsilon^2\calI)\). Since the observation operator is the identity and \(\calC=S^\top\Lambda S\), the posterior is \(\mu_1=\calN(m_\star,\calC_\star)\), where \(\calC_\star=(\calC^{-1}+\sigma_\epsilon^{-2}\calI)^{-1}\), \(m_\star=\sigma_\epsilon^{-2}\calC_\star y\), and \(\calC_\star=S^\top\Lambda_\star S\) with \((\Lambda_\star)_{kk}=\sigma_k/(1+\sigma_\epsilon^{-2}\sigma_k)\). Thus the KL-optimal parameters are available exactly: the optimal mean is \(m_\star\), the optimal variance multipliers are \(d_{\star,k}=(1+\sigma_\epsilon^{-2}\sigma_k)^{-1}\), and \(\mu_1=\mu_{d_\star,m_\star}\), so \(\widetilde\Phi_{d_\star,m_\star}\) is constant.

\begin{figure*}[t]
    \centering
    \begin{subfigure}[t]{0.48\textwidth}
        \centering
        \includegraphics[width=\linewidth]{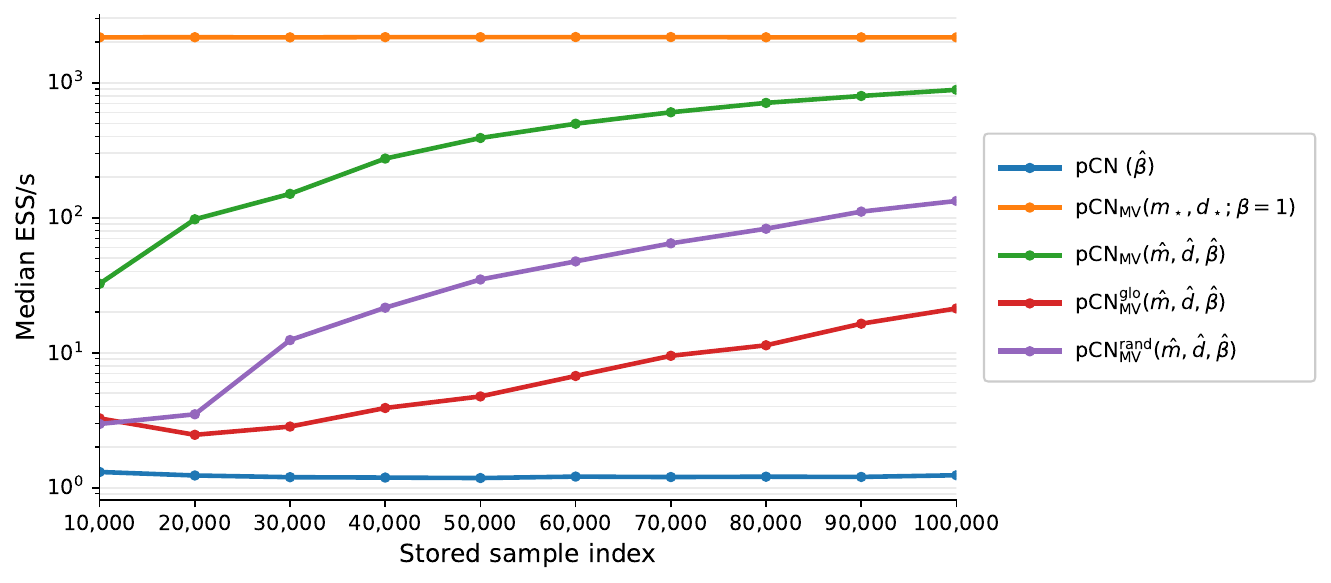}
        \caption{\(\mathrm{pCN}_{\mathrm{MV}}\).}
        \label{fig:pcnmv-ess}
    \end{subfigure}
    \hfill
    \begin{subfigure}[t]{0.48\textwidth}
        \centering
        \includegraphics[width=\linewidth]{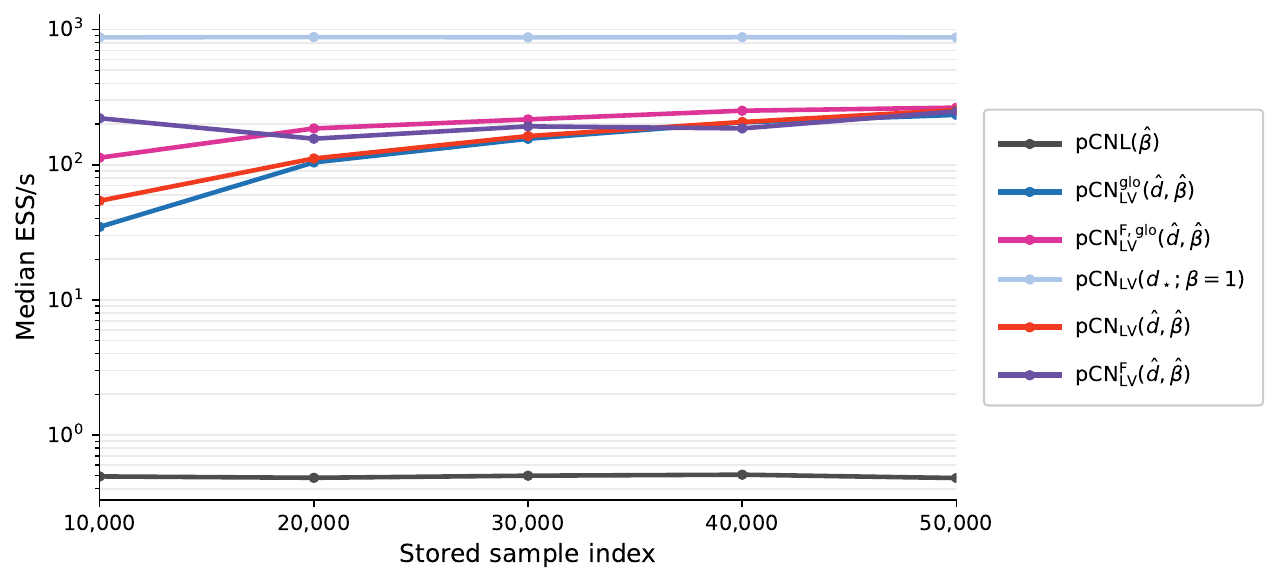}
        \caption{\(\mathrm{pCNL}_{\mathrm V}\).}
        \label{fig:pcnlv-ess}
    \end{subfigure}
    \caption{Gaussian benchmark: median ESS/s over consecutive chunks of \(10^4\) stored samples. The oracle sampler gives an upper benchmark, while the adaptive truncated schemes substantially improve over the corresponding baselines.}
    \label{fig:gaussian-ess}
\end{figure*}

We discretize \([0,1]\) using \(400\) equally spaced grid points and use a Mat\'ern Gaussian process prior \citep{matern1960} with marginal variance \(1.0\), smoothness \(\nu=1.5\), inverse length-scale \(\kappa=1.5\), and noise variance \(\sigma_\epsilon=0.25\). For \(\text{pCN}_{\text{MV}}\), we compare a truncated cyclic update, a truncated random-scan update, and a global update without truncation, with the oracle sampler \(\text{pCN}_{\text{MV}}(m_\star,d_\star;\beta=1)\) serving as an upper reference. Figure~\ref{fig:gaussian-ess} reports median ESS/s over chunks of \(10^4\) stored samples. The truncated online \(\text{pCN}_{\text{MV}}(\hat m,\hat d,\hat\beta)\) scheme rapidly improves over \(\text{pCN}(\hat\beta)\) and eventually achieves orders-of-magnitude larger ESS/s; the global and random-scan variants also improve over the baseline, but less strongly. For the \(\text{pCNL}_{\text{V}}\) variants, the ordering of the adaptation is less critical, suggesting that errors in \(\hat d\) can be partly compensated by adapting \(\beta\).

We close with a diagnostic illustrating why online truncation is important.
In Figure~\ref{fig:pcnmv-adapt} the
left panel shows that the truncated estimator stabilizes the leading variance
multipliers \(d_k\) quickly, while the global update remains noisy in the
leading coordinates. The middle panel shows the resulting \(\beta\)-adaptation
for \(\mathrm{pCN}_{\mathrm{MV}}\): the truncated scheme reaches
\(\beta\approx1\) after roughly \(2\times10^4\) iterations, whereas the global
update has not converged after \(10^5\) iterations. The right panel shows the
corresponding \(\beta\)-adaptation for \(\mathrm{pCNL}_{\mathrm V}\), where the
Fisher-scaled adaptation converges substantially faster. Overall, these
diagnostics show that finite active adaptation can be much more effective than
adapting all coordinates simultaneously, and that Fisher scaling can further
accelerate gradient-based adaptation during the inital phase.


\subsection{A Darcy flow inverse problem with a DNA prior}
\label{subsec:darcy-dna}

We next test the proposed adaptation in a nonlinear PDE-constrained inverse problem, where the posterior is no longer Gaussian and the optimal change of measure is not available in closed form. We consider the Darcy flow model studied by \citet{wang2026global}. On \(D=[0,1]^2\), the unknown is a positive conductivity field \(f\), and the forward map \(G(f)\) returns the solution \(u\) of
$
\nabla\cdot(f\nabla u)=g \quad \text{in }D,$ where
$u=0 \quad \text{on }\partial D $. The data are noisy point observations,
$
Y_i=G(f_0)(x_i)+\epsilon_i,
$
where $
\epsilon_i\sim\calN(0,0.01^2),
$
at \(500\) randomly selected finite-volume cell centers. We infer \(f_0\) through the log-conductivity parameterization \(f=\exp(u)\), with a Mat\'ern Gaussian prior on \(u\). The prior is represented using the Dirichlet--Neumann averaging (DNA) construction of \citet{kutri2026dna}; full discretization, prior, and solver details are given in Appendix~\ref{app:darcy-diagnostics}.

The DNA construction is particularly well suited to the proposed methods, since it provides efficient application of the transformations \(S\) and \(S^{-1}\), while the diagonal operator \(\Lambda\) is available in closed form. These properties make it straightforward to implement the adapted Gaussian-reference proposals without forming dense covariance matrices.

For this nonlinear Darcy problem, however, the gradient of the log-likelihood is not available in closed form, and each likelihood evaluation requires solving the discretized elliptic PDE. Since this makes gradient-based, and more extensive algorithmic, comparisons computationally expensive, we restrict the experiment to a direct comparison between the baseline \(\text{pCN}\) sampler and the proposed \(\text{pCN}_{\text{MV}}\) method.

On the sampled log-likelihood trace with \(48000\) post-burn-in samples after
\(2000\) burn-in iterations, adaptive \(\text{pCN}\) gives ESS \(1747.6\),
whereas \(\text{pCN}_{\text{MV}}\) gives ESS \(7742.5\), corresponding to
more than a four-fold improvement. The estimated posterior mean recovers the
dominant spatial structure of the conductivity field; the field comparison,
trace plots, and adaptation diagnostics are deferred to
Appendix~\ref{app:darcy-diagnostics}.
\subsection{Bayesian logistic-regression benchmarks}
\label{subsec:logistic-regression}

We finally consider Bayesian binary logistic regression on five standard
datasets, following the setup of \citet{titsias2016auxiliary,titsias2023fisher}.
For observations \(\{(y_i,z_i)\}_{i=1}^n\), with \(y_i\in\{0,1\}\), we use a
sigmoid likelihood and a standard Gaussian prior over the latent function values.
The covariates are centered and scaled, and then used as input locations for the
Gaussian-process prior. The resulting latent state dimension is the number of
observations, and $p$ is the number of covariates: Australian \((n=690,p=14)\), Heart \((n=270,p=13)\), German
\((n=1000,p=24)\), Pima \((n=532,p=7)\), and Ripley \((n=250,p=2)\).   Additional implementation details and diagnostics are given in
Appendix~\ref{app:logistic-regression}.

We compare \(\text{pCN}\), \(\text{pCNL}\), the variance-adapted
\(\text{pCNL}_{\text{V}}\), its Fisher-scaled variant
\(\text{pCNL}_{\text{V}}\)-F, the mean--variance-adapted
\(\text{pCN}_{\text{MV}}\), and the auxiliary marginal sampler of
\citet{titsias2016auxiliary}. The truncated adaptive schemes start with
\(N_0=20\) active coordinates, add \(5\) coordinates at each growth step, and
grow every \(1000\) iterations. Table~\ref{tab:gpcreate-logit-state-ess-pivot}
reports median ESS and median ESS/s for the latent Gaussian-process state. The
adaptive methods substantially improve over pCN and pCNL on all datasets,
showing that learning posterior scale in the leading prior directions can be
beneficial even in finite-dimensional non-Gaussian classification problems.

\begin{table}[!htbp]
\caption{Bayesian logistic regression: median ESS and median ESS/s in state
space. Values are ESS (ESS/s); bold marks the highest ESS/s per dataset.}
\label{tab:gpcreate-logit-state-ess-pivot}
\centering
\small
\begin{tabular}{lrrrrr}
\toprule
Method & Australian & Heart & German & Pima & Ripley \\
\midrule
pCN & 524 (2.92) & 1313 (32.3) & 460 (1.19) & 2473 (18.6) & 1359 (40.6) \\
pCNL & 754 (1.48) & 3712 (37.4) & 1157 (1.15) & 6854 (19.2) & 2806 (31.1) \\
pCNLV & 46065 (82.4) & 53871 (472) & 59799 (51.5) & 113523 (302) & \textbf{18876 (201)} \\
pCNLV-F & 45319 (76.1) & \textbf{54236 (479)} & \textbf{60119 (59.3)} & 122810 (348) & 18894 (182) \\
pCNMV & \textbf{17423 (88.2)} & 21088 (422) & 10114 (26.8) & \textbf{82365 (502)} & 6187 (148) \\
AuxMarg & 28634 (50.9) & 35307 (329) & 32316 (27.3) & 68879 (202) & 15708 (195) \\
\bottomrule
\end{tabular}
\end{table}


\bibliographystyle{plainnat}
\bibliography{aMCMCinf}


\appendix

\section{Additional diagnostics for the Gaussian benchmark}
\label{app:gaussian-benchmark}
The Gaussian benchmark in Section~\ref{subsec:gaussian-benchmark} was run on a MacBook Pro (\texttt{MacBookPro16,1}) with one 8-core CPU at \(2.3\) GHz, 16 GB memory, macOS 14.5, and Python 3.12.12 from \path{/Users/jonaswallin/Library/r-miniconda/bin/python3.12}. 

The experiment  uses \(400\) uniformly spaced grid points on \([0,1]\), and sets the GPCreate Mat\'ern prior parameters to variance \(1.0\), \(\kappa=1.5\), \(\nu=1.5\), and jitter \(10^{-10}\). The observation operator is the identity matrix, so every grid point is observed once. Data are generated by drawing \(z_{\star}\) from the prior, mapping it to state space, and adding independent Gaussian measurement noise with standard deviation \(0.25\). Because the model is linear Gaussian, the posterior mean and posterior variance multipliers are computed analytically and are used both as an oracle reference and as the target for the online adaptation diagnostics.

For the \(\mathrm{pCN}_{\mathrm{MV}}\) comparison,  used \(102000\) MCMC iterations, discards \(2000\) iterations as burn-in, and computes chunk diagnostics using blocks of \(10000\) stored samples. The beta-only \(\mathrm{pCN}\) chain starts from \(\beta=0.03\) and targets acceptance probability \(0.234\). The adaptive \(\mathrm{pCN}_{\mathrm{MV}}\) chains use the same initial \(\beta\), beta step size \(0.5\), and beta updates every \(25\) iterations. The truncated online chain starts with \(N_0=16\) active coordinates, adds \(16\) coordinates every \(1000\) iterations, and orders coordinates by decreasing prior variance. The global comparison updates all coordinates from the beginning, while the random-order comparison uses the same truncation schedule but a random coordinate order with seed \(42\).

\begin{figure}[t]
    \centering
    \begin{subfigure}[t]{0.48\linewidth}
        \centering
        \includegraphics[width=\linewidth]{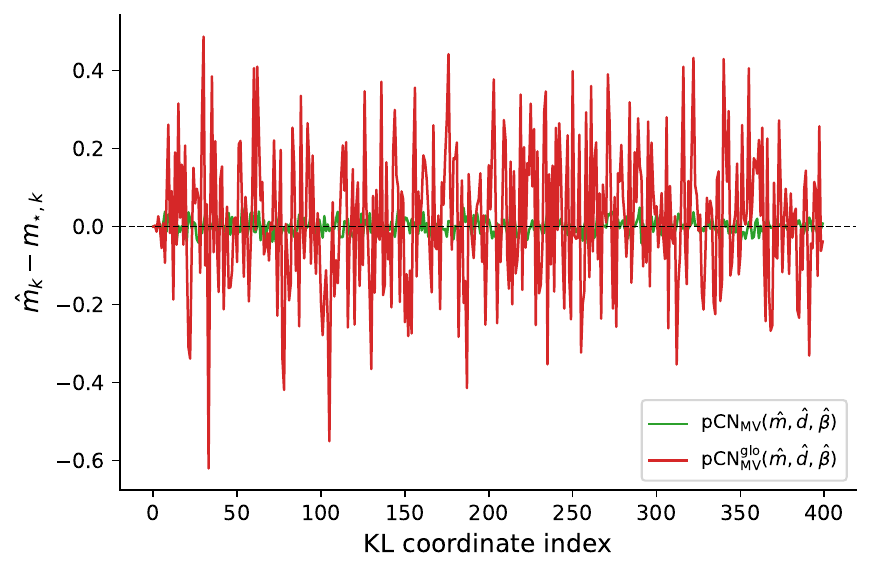}
        \caption{KL-coordinate discrepancy in the mean adaptation.}
        \label{fig:pcnmv-m-app}
    \end{subfigure}\hfill
    \begin{subfigure}[t]{0.48\linewidth}
        \centering
        \includegraphics[width=\linewidth]{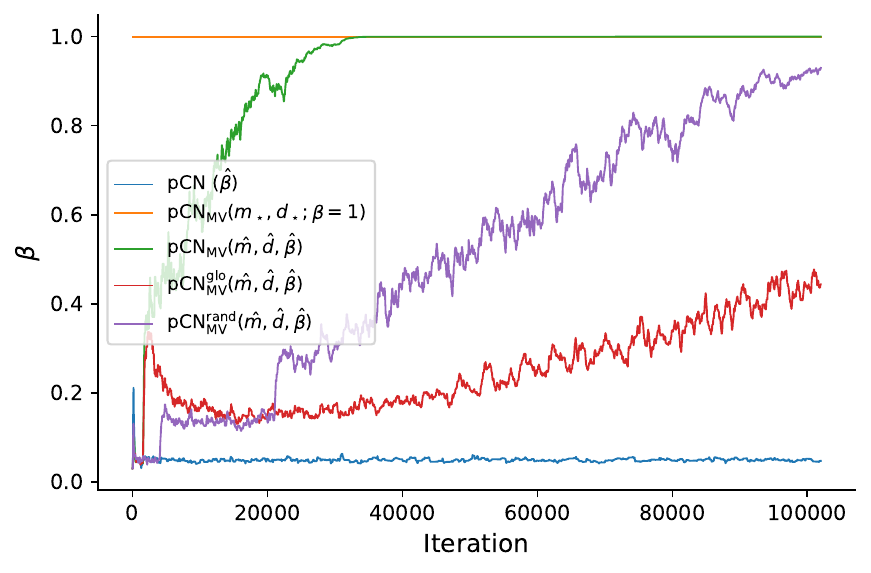}
        \caption{Trace of the proposal scale \(\beta\).}
        \label{fig:pcnmv-beta-app}
    \end{subfigure}
    \caption{Additional diagnostics for the Gaussian benchmark. The left panel shows the KL-coordinate discrepancy in the estimated mean adaptation, while the right panel shows the evolution of the proposal scale \(\beta\) across samplers.}
    \label{fig:pcnmv-diagnostics-app}
\end{figure}

Figure~\ref{fig:pcnmv-diagnostics-app} provides additional diagnostics for the Gaussian benchmark. The truncated online scheme gives a more accurate estimate of the leading mean-adaptation coordinates, and its proposal scale \(\beta\) evolves toward a substantially more effective regime than the baseline \(\mathrm{pCN}(\hat\beta)\).
\section{Additional details and diagnostics for the Darcy flow example}
\label{app:darcy-diagnostics}

This appendix gives the implementation details and additional diagnostics for the Darcy flow experiment in Section~\ref{subsec:darcy-dna}. The data are noisy point observations
\[
Y_i = G(f_0)(x_i)+\epsilon_i,
\qquad
\epsilon_i\sim \calN(0,\sigma_\epsilon^2),
\]
at randomly selected finite-volume cell centers. The experiment was run on the MacBook Pro described in Appendix~\ref{app:gaussian-benchmark}. The experiment uses \(500\) observations, and sets \(\sigma_\epsilon=0.01\). The log-conductivity prior has Mat\'ern parameters \((\ell,\nu,\mathrm{var})=(0.3,2.1,0.35)\). The DNA parameter is \(q=32\), giving \(4225\) latent coefficients and a \(34\times34\) finite-volume grid. The source \(g\) is fixed across samplers after drawing an independent DNA Mat\'ern field with parameters \((0.2,5.1,0.12)\) and exponentiating it. The finite-volume solves are computed with FiPy \citep{guyer2009fipy}.

The synthetic truth is generated by drawing \(z_{f_0}\) and \(z_g\) independently from their DNA Mat\'ern priors, transforming them to grids, exponentiating to obtain \(f_0\) and \(g\), and solving the elliptic equation once to obtain \(G(f_0)\). Observation locations are sampled without replacement from the finite-volume cell centers, and the observed values are \(G(f_0)\) at those locations plus independent Gaussian noise. The sampler starts at the zero coefficient vector. The likelihood object uses a forward finite-difference gradient on the \(8\) largest-prior-variance DNA coordinates, with finite-difference step \(10^{-4}\) and cache size \(256\); this gradient setting is present for gradient-based variants, although the reported comparison uses pCN and \(\mathrm{pCN}_{\mathrm{MV}}\).

We compare the beta-adapted pCN baseline with the online truncated \(\mathrm{pCN}_{\mathrm{MV}}(\hat m,\hat d,\hat\beta)\) sampler. Both chains use \(50000\) MCMC iterations and discard \(2000\) iterations as burn-in. The adaptive pCN baseline starts from \(\beta=0.05\) and targets acceptance probability \(0.2\). The \(\mathrm{pCN}_{\mathrm{MV}}\) chain also starts from \(\beta=0.05\), targets acceptance probability \(0.234\), starts with \(N_0=16\) active coordinates, adds \(16\) coordinates every \(1000\) iterations, initializes \(d_k=1\), and initializes the second-moment count at \(2\). Log-likelihood ESS diagnostics are computed from the stored post-burn-in log-likelihood trace, and the running ESS plot uses chunks of \(10000\) stored samples. 

\begin{figure*}[t]
    \centering
    \begin{subfigure}[t]{0.32\linewidth}
        \centering
        \includegraphics[width=\linewidth]{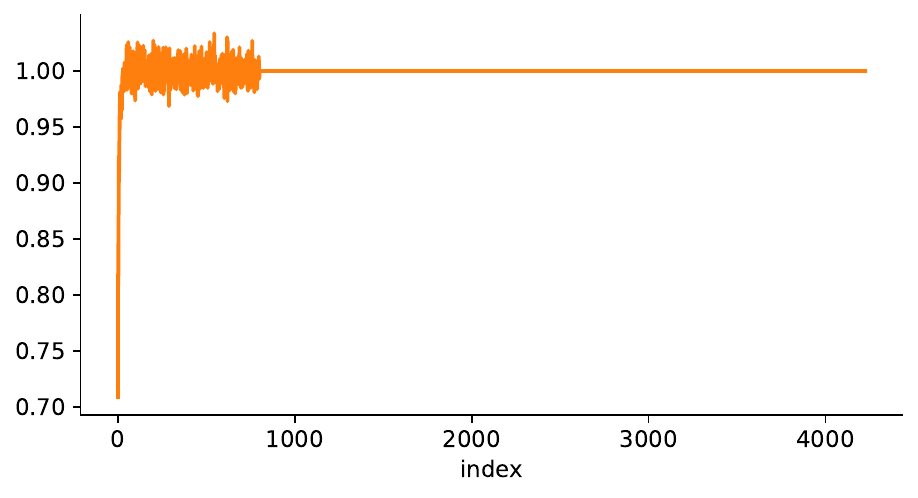}
        \caption{Final ordered \(\hat d\)-estimate.}
        \label{fig:darcy-d-app}
    \end{subfigure}\hfill
    \begin{subfigure}[t]{0.32\linewidth}
        \centering
        \includegraphics[width=\linewidth]{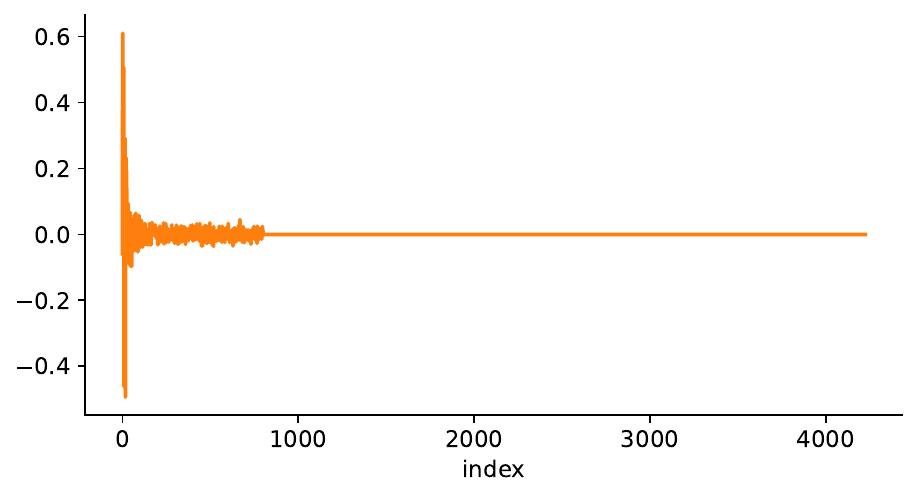}
        \caption{Final ordered \(\hat m\)-estimate.}
        \label{fig:darcy-m-app}
    \end{subfigure}\hfill
    \begin{subfigure}[t]{0.32\linewidth}
        \centering
        \includegraphics[width=\linewidth]{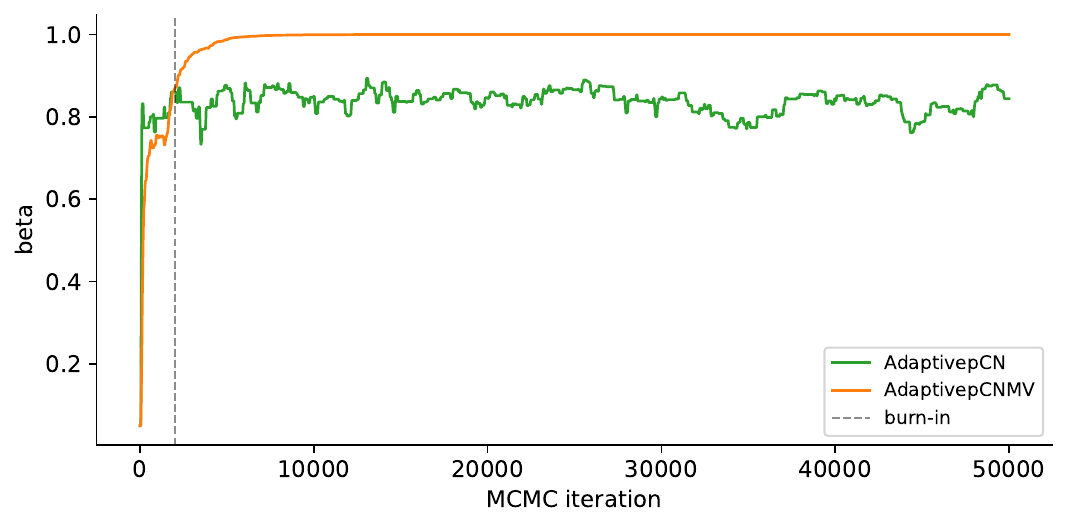}
        \caption{Adapted proposal scale \(\beta\).}
        \label{fig:darcy-beta-app}
    \end{subfigure}
    \caption{Adaptation diagnostics for the Darcy flow experiment. The first two panels show the final truncated online estimates used by \(\mathrm{pCN}_{\mathrm{MV}}\); the last panel compares proposal-scale adaptation for adaptive pCN and \(\mathrm{pCN}_{\mathrm{MV}}\).}
    \label{fig:darcy-adaptation-app}
\end{figure*}

\begin{figure*}[t]
    \centering
    \begin{subfigure}[t]{0.48\linewidth}
        \centering
        \includegraphics[width=\linewidth]{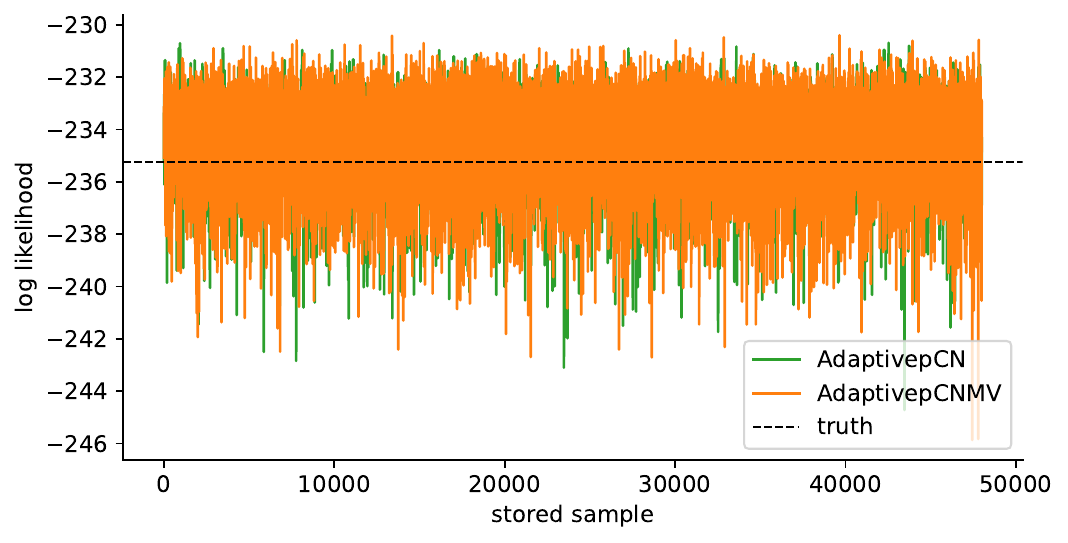}
        \caption{Stored log-likelihood traces.}
        \label{fig:darcy-loglik-trace-app}
    \end{subfigure}\hfill
    \begin{subfigure}[t]{0.48\linewidth}
        \centering
        \includegraphics[width=\linewidth]{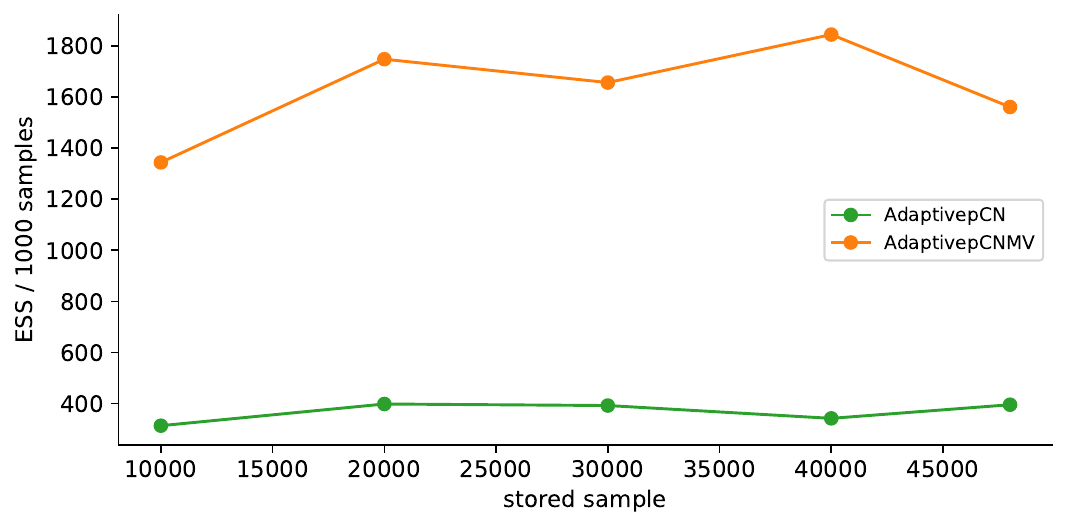}
        \caption{Running likelihood ESS per \(1000\) stored samples.}
        \label{fig:darcy-ess-per-1000-app}
    \end{subfigure}
    \caption{Likelihood diagnostics for the Darcy flow experiment.}
    \label{fig:darcy-likelihood-app}
\end{figure*}

Figure~\ref{fig:darcy-adaptation-app} shows the final online adaptation estimates and the proposal-scale trace. The ordered \(\hat d\)-estimate shows the learned variance rescaling in the truncated coefficient order, while the ordered \(\hat m\)-estimate shows the learned posterior mean shift used by \(\mathrm{pCN}_{\mathrm{MV}}\). Figure~\ref{fig:darcy-likelihood-app} shows the stored log-likelihood traces and the running likelihood ESS per \(1000\) stored samples.

\section{Additional details for the Bayesian logistic-regression experiments}
\label{app:logistic-regression}

This appendix gives implementation details and additional diagnostics for the
Bayesian logistic-regression experiments reported in
Section~\ref{subsec:logistic-regression}. We consider five standard binary
classification datasets, following the setup of
\citet{titsias2016auxiliary,titsias2023fisher}. For labels
\(y_i\in\{-1,1\}\), the latent Gaussian-process values
\(u=(u_1,\ldots,u_n)\) enter the likelihood as
\[
p(Y\mid u)=\prod_{i=1}^n \sigma(y_i u_i),
\]
where \(\sigma(t)=(1+\exp(-t))^{-1}\). The original binary labels are therefore
mapped to \(\{-1,1\}\) before sampling.

Each dataset is stored as a matrix whose final column contains the binary
response. The covariates are centered and scaled to unit empirical standard
deviation, and then used as input locations for the Gaussian-process prior. We
use the GPCreate construction at the observed covariate locations; the
dataset-specific prior variance and length-scale are fixed before sampling. The
resulting latent state dimension is the number of observations in the
corresponding dataset: Australian \((n=690,p=14)\), Heart
\((n=270,p=13)\), German \((n=1000,p=24)\), Pima \((n=532,p=7)\), and Ripley
\((n=250,p=2)\).

For each dataset, we run six samplers: \(\mathrm{pCN}\), \(\mathrm{pCNL}\),
\(\mathrm{pCNL}_{\mathrm V}\), \(\mathrm{pCNL}_{\mathrm V}\)-F,
\(\mathrm{pCN}_{\mathrm{MV}}\), and the auxiliary marginal sampler of
\citet{titsias2016auxiliary}. The initial proposal scales are \(0.08\) for
\(\mathrm{pCN}\) and \(\mathrm{pCNL}\), \(0.2\) for
\(\mathrm{pCNL}_{\mathrm V}\) and \(\mathrm{pCNL}_{\mathrm V}\)-F, and \(0.3\)
for \(\mathrm{pCN}_{\mathrm{MV}}\). The proposal scale is adapted during
sampling: \(\mathrm{pCN}\) and \(\mathrm{pCN}_{\mathrm{MV}}\) target average
acceptance probability \(0.234\), while the gradient-based schemes target
\(0.54\). The truncated adaptive schemes start with \(N_0=20\) active
coordinates, add \(5\) coordinates at each growth step, and grow every \(1000\)
iterations.

\begin{figure*}[t]
    \centering
    \includegraphics[width=0.72\linewidth]{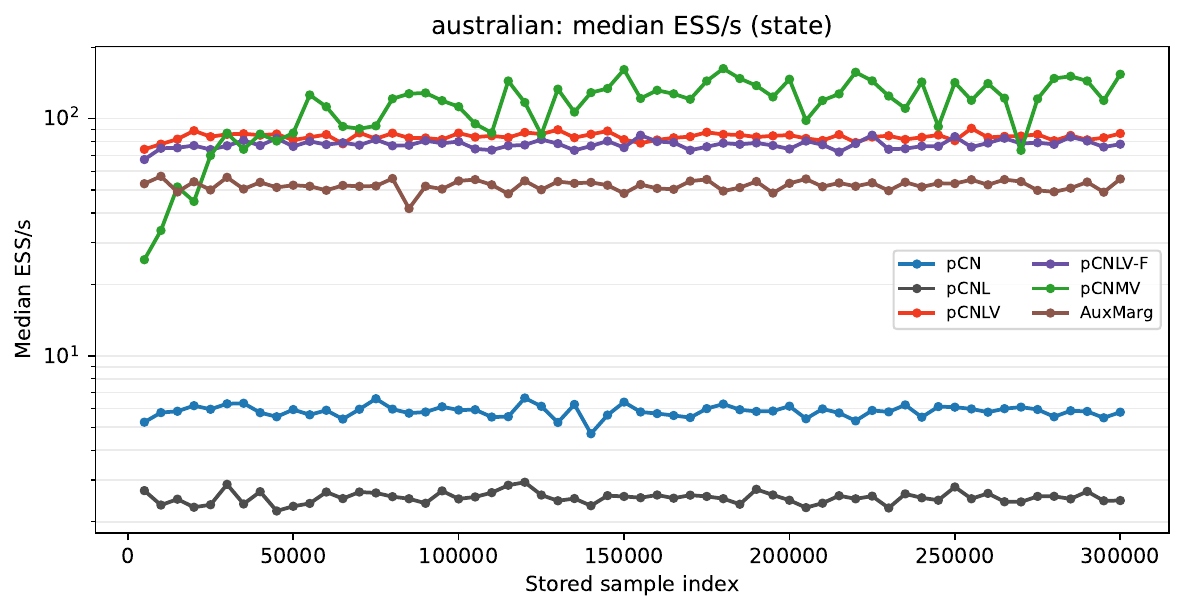}
    \caption{Australian dataset: median state-space ESS/s over consecutive chunks of \(5000\) stored samples.}
    \label{fig:gpcreate-logit-australian-chunks}
\end{figure*}

\begin{figure*}[t]
    \centering
    \includegraphics[width=0.72\linewidth]{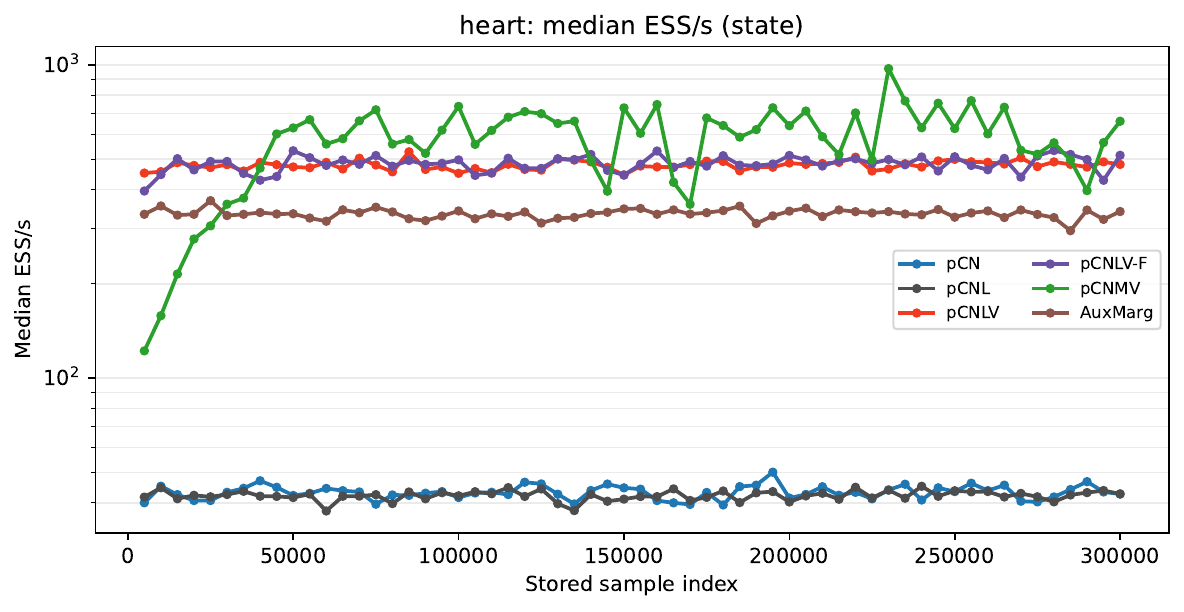}
    \caption{Heart dataset: median state-space ESS/s over consecutive chunks of \(5000\) stored samples.}
    \label{fig:gpcreate-logit-heart-chunks}
\end{figure*}

\begin{figure*}[t]
    \centering
    \includegraphics[width=0.72\linewidth]{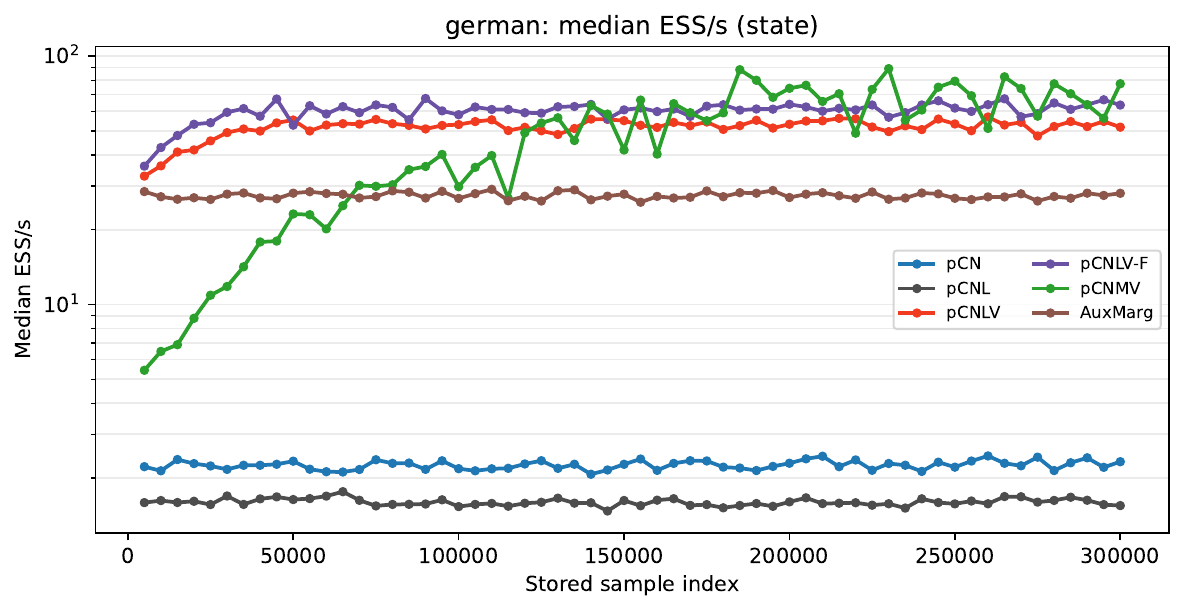}
    \caption{German dataset: median state-space ESS/s over consecutive chunks of \(5000\) stored samples.}
    \label{fig:gpcreate-logit-german-chunks}
\end{figure*}

\begin{figure*}[t]
    \centering
    \includegraphics[width=0.72\linewidth]{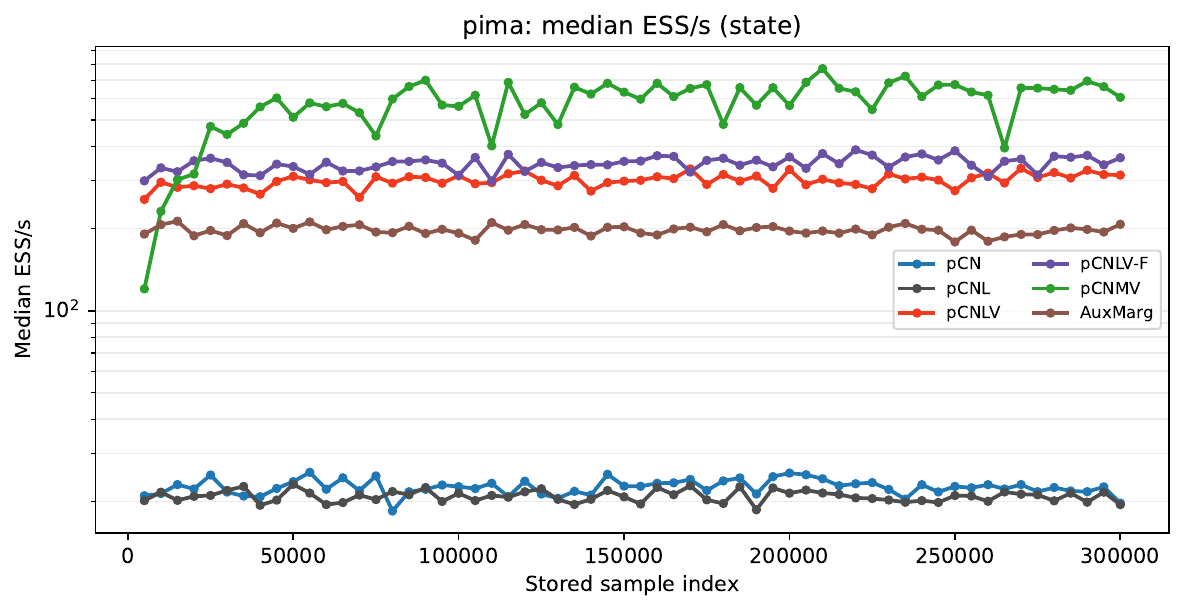}
    \caption{Pima dataset: median state-space ESS/s over consecutive chunks of \(5000\) stored samples.}
    \label{fig:gpcreate-logit-pima-chunks}
\end{figure*}

\begin{figure*}[t]
    \centering
    \includegraphics[width=0.72\linewidth]{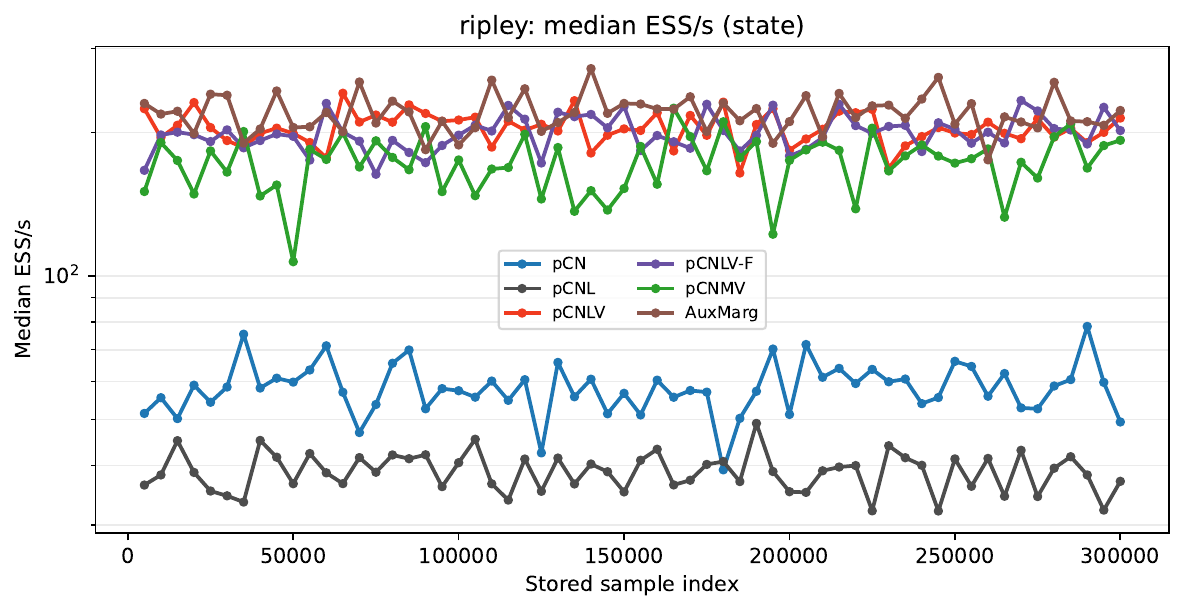}
    \caption{Ripley dataset: median state-space ESS/s over consecutive chunks of \(5000\) stored samples.}
    \label{fig:gpcreate-logit-ripley-chunks}
\end{figure*}

Each dataset--method pair is run for \(310000\) MCMC iterations. We discard the
first \(10000\) iterations as burn-in, store every remaining draw, and compute
chunk diagnostics on consecutive blocks of \(5000\) stored samples. The
aggregate median ESS and median ESS/s values are reported in
Table~\ref{tab:gpcreate-logit-state-ess-pivot}; Figures
\ref{fig:gpcreate-logit-australian-chunks}--\ref{fig:gpcreate-logit-ripley-chunks}
show the corresponding evolution of median state-space ESS/s across chunks.

\begin{table*}[t]
\caption{AWS ParallelCluster configuration used for the logistic-regression sampling runs.}
    \label{tab:gpcreate-logit-cluster}
    \centering
    \small
    \begin{tabular}{lp{0.70\linewidth}}
        \toprule
        Item & Setting \\
        \midrule
        AWS profile & \texttt{pcluster} AWS CLI profile. \\
        Cluster manager and scheduler & AWS ParallelCluster with Slurm, queue \texttt{queue1}. \\
        Operating system image & Ubuntu 22.04. \\
        Compute nodes & 30 \texttt{m6i.large} instances. Each node provides 2 vCPU and 8 GiB memory. \\
        Slurm job layout & One array job with 30 tasks, corresponding to \(5\) datasets times \(6\) methods. Each task requests one node, one CPU, and exclusive node access. \\
        Python environment & Head-node virtual environment with Python 3.10, NumPy 2.2.6, SciPy 1.15.3, and Matplotlib 3.10.9. \\
        \bottomrule
    \end{tabular}
\end{table*}

\begin{table*}[t]
    \caption{Final adapted proposal scale \(\beta\) for each logistic-regression dataset and sampler. Values are rounded to three decimal places.}
    \label{tab:gpcreate-logit-beta}
    \centering
    \small
    \begin{tabular}{lrrrrr}
        \toprule
        Method & Australian & Heart & German & Pima & Ripley \\
        \midrule
        \(\mathrm{pCN}\) & 0.120 & 0.198 & 0.118 & 0.259 & 0.144 \\
        \(\mathrm{pCNL}\) & 0.093 & 0.215 & 0.126 & 0.268 & 0.138 \\
        \(\mathrm{pCNL}_{\mathrm V}\) & 0.900 & 0.933 & 0.953 & 1.000 & 0.605 \\
        \(\mathrm{pCNL}_{\mathrm V}\)-F & 0.937 & 0.957 & 0.969 & 1.000 & 0.869 \\
        \(\mathrm{pCN}_{\mathrm{MV}}\) & 1.000 & 1.000 & 1.000 & 1.000 & 0.617 \\
        AuxMarg & 0.968 & 0.950 & 0.896 & 0.967 & 0.983 \\
        \bottomrule
    \end{tabular}
\end{table*}

\section{Online estimation of the Fisher multipliers.}
\label{sec:ONline}
A simple online estimator corresponding to \eqref{eqn:fisher-dk} is based on the whitened score vector
\[
g^{(j)}
:=
\Lambda^{1/2}S \Bigl(\grad\widetilde\Phi_{d,0}(u^{(j)}) + \calC_d^{-1} u^{(j)} \Bigr).
\]
Note that, by the definition of \(\widetilde\Phi_{d,0}\),
$
\grad\widetilde\Phi_{d,0}(u) + \calC_d^{-1}u
=
\grad\Phi(u)+\calC^{-1}u,
$
so \(g^{(j)}\) is the whitened score vector of the target in the prior Karhunen--Lo\`eve coordinates. Writing \(g_k^{(j)}\) for its \(k\)-th component, we estimate its first and second moments recursively by
\begin{align*}
\hat r_k^{(j)}
&=
w_j g_k^{(j)} + (1-w_j)\hat r_k^{(j-1)},\\
\hat q_k^{(j)}
&=
w_j \bigl(g_k^{(j)}\bigr)^2 + (1-w_j)\hat q_k^{(j-1)},
\end{align*}
with \(w_j=j^{-1}\). We then define the centered Fisher estimator
$
\hat\iota_k^{(j)}
:=
\hat q_k^{(j)}-\bigl(\hat r_k^{(j)}\bigr)^2.
$
This uses the empirical covariance of the score components rather than their raw second moment, which is more stable during the burn-in phase when the chain is not yet close to stationarity and \(\E[g_k]\) need not be negligible.

We then set
$\hat d_{k,\mathrm F}^{(j)}
=
\bigl(\hat\iota_k^{(j)}\bigr)^{-1}$. To preserve equivalence with \(\mu_0\), we use the same truncated adaptation strategy as above and define
\[
\tilde d_{k,\mathrm F}^{(j)}
=
\begin{cases}
\hat d_{k,\mathrm F}^{(j)}, & k < N_j,\\
1, & k \ge N_j.
\end{cases}
\]
By construction, \(\tilde d_{\mathrm F}^{(j)}-1\) has finite support, so
\(\sum_{k\ge1}(\tilde d_{k,\mathrm F}^{(j)}-1)^2<\infty\) holds automatically at every iteration.


\section{Proofs}
\label{sec:proof}
\begin{proof}[Proof of Lemma \ref{lem:1}]
Let \(\mathcal F_{r-1}\) denote the history before the \(r\)-th proposal.
Conditional on \(\mathcal F_{r-1}\), write the proposal as
\[
v^{(r)}=c_rh^{(r)}+\calC_{d^{(r)}}^{1/2}\eta^{(r)},
\]
where \(h^{(r)}\) is \(\mathcal F_{r-1}\)-measurable,
\(\eta^{(r)}\sim\calN(0,\calI)\), and \(\eta^{(r)}\) is independent of
\(\mathcal F_{r-1}\). In \(\calC\)-whitened coordinates,
\[
\calC^{-1/2}v^{(r)}
=
\calC^{-1/2}(c_rh^{(r)})+D_r^{1/2}\eta^{(r)},
\qquad
D_r=\operatorname{diag}(d_k^{(r)}).
\]
Under the truncated adaptation, \(d_k^{(r)}=1\) for all sufficiently large
\(k\). Thus the tail coordinates have the form
\(\rho_k+\eta_k^{(r)}\), where \((\rho_k)_{k\ge1}\) is fixed conditional on
\(\mathcal F_{r-1}\). More generally, for any
\(\mathcal F_{r-1}\)-measurable sequence \((\rho_k)_{k\ge1}\) and any
\(a\neq0\),
\[
\sum_{k\ge1}(\rho_k+a\eta_k^{(r)})^2=\infty
\qquad\text{a.s.}
\]
Indeed, conditional on \(\mathcal F_{r-1}\), the events
\(\{|\rho_k+a\eta_k^{(r)}|>|a|/2\}\) are independent and have probabilities
bounded below by a positive constant, uniformly in \(\rho_k\). Hence infinitely
many occur by the Borel--Cantelli lemma. Therefore
\[
\mathbb P\bigl(v^{(r)}\in\calH_{\calC}\mid\mathcal F_{r-1}\bigr)=0 .
\]

Since \(u^{(0)}\sim\mu_1\), \(\mu_1\ll\mu_0\), and
\(\mu_0(\calH_{\calC})=0\), we have \(u^{(0)}\notin\calH_{\calC}\) almost
surely. Each Metropolis--Hastings update sets \(u^{(r)}\) equal either to
\(u^{(r-1)}\) or to \(v^{(r)}\). Hence, by induction,
\(u^{(r)}\notin\calH_{\calC}\) almost surely for every \(r\ge0\).

Now fix \(j\). If no proposal has been accepted up to time \(j\), then all
states in the average equal \(u^{(0)}\), so
\(\hat m_{\mathrm{orig}}^{(j)}=u^{(0)}\notin\calH_{\calC}\) almost surely.
Otherwise, let \(k\le j\) be the last accepted proposal time. Then
\(u^{(k)}=\cdots=u^{(j)}=v^{(k)}\). Consequently, in \(\calC\)-whitened
coordinates,
\[
\calC^{-1/2}\hat m_{\mathrm{orig}}^{(j)}
=
\calC^{-1/2}\rho+c\,\calC^{-1/2}v^{(k)}
=
\rho'+cD_k^{1/2}\eta^{(k)},
\qquad c>0,
\]
where \(\rho\) and \(\rho'\) are \(\mathcal F_{k-1}\)-measurable coordinate
sequences. By the preceding
Borel--Cantelli argument, 
\(\hat m_{\mathrm{orig}}^{(j)}\notin\calH_{\calC}\).

For the truncated estimator, \(\tilde m_k^{(j)}=0\) for all \(k\ge N_j\). Thus
\(\tilde m_{\mathrm{orig}}^{(j)}=S^\top\Lambda\tilde m^{(j)}\) has only
finitely many nonzero coordinates in the \(\calC\)-eigenbasis, and therefore
\(\tilde m_{\mathrm{orig}}^{(j)}\in\calH_{\calC}\) almost surely.

Finally, the active variance multipliers are strictly positive and
\(\tilde d^{(j)}-1\) has finite support by construction. Since the truncated mean belongs to
\(\calH_{\calC}\), the Feldman--Hajek criterion implies that the adapted
Gaussian reference remains equivalent to the prior at every iteration.
\end{proof}

\end{document}